\documentclass[twocolumn,showpacs,superscriptaddress,prb,amsmath,amssymb,floatfix]{revtex4}

\usepackage{amsmath,amssymb}
\usepackage{bm}
\usepackage{epsfig}
\usepackage{psfrag}

\begin{document}
 \title{Self-consistent Fermi surface renormalization of two coupled Luttinger liquids}
  
 
\author{Sascha Ledowski}
 \affiliation{Institut f\"{u}r Theoretische Physik, Universit\"{a}t
    Frankfurt,  Max von Laue-Str. 1, 60438 Frankfurt, Germany}

\author{Peter Kopietz}

\author{Alvaro Ferraz}
\affiliation{
International Center for Condensed Matter Physics,
Universidade de  Bras{\'\i}{}lia, 70910-900   Bras{\'\i}{}lia, Brazil}

%

\date{December 22, 2004}

  \begin{abstract}
 Using functional renormalization group methods, we present a
self-consistent calculation of
the true Fermi momenta $k_F^a$ (antibonding band) 
and $k_F^b$ (bonding band) of two spinless interacting
metallic chains coupled by 
small interchain  hopping.
In the regime where the system is a Luttinger liquid,
we find that
$\Delta = k_F^b - k_F^a$ 
is self-consistently determined by 
$\Delta = \Delta_{1} [ 1 +   {g}_0^2 \ln ( \Lambda_0  / \Delta )^2]^{-1}$,
where ${g}_0$ is  the dimensionless
interchain backscattering interaction,
$\Delta_{1}$ is the Hartree-Fock result for
$k_F^{b}-k_F^a$,
and 
$\Lambda_0 \gg \Delta $ is an ultraviolet cutoff.
For $ {g}_0^2 \ln ( \Lambda_0 / \Delta_{1} )^2 
\gg 1$ even weak interachain backscattering leads to a strong
reduction of the distance between the Fermi momenta.

  \end{abstract}




  \maketitle
 
\section{Introduction}

What survives of the Luttinger liquid properties of a one-dimensional metal
if two or more metallic chains are coupled by some weak
interchain hopping $t_{\bot}$?  
Motivated by the proposal 
that the anomalous normal state
properties of the high-temperature superconductors 
are a manifestation of a two-dimensional non-Fermi liquid state~\cite{Anderson87},
many authors tried to find an answer to the above question, using
non-perturbative methods such as bosonization or the   
renormalization group (RG) 
\cite{Bourbonnais91,Kusmartsev92,Finkelstein93,Fabrizio93,Balents96,Arrigoni98,Ledermann00,Louis01,Dusuel03}.
As a first step
towards a solution of this rather difficult 
problem, it is instructive to study
just two coupled spinless metallic chains~\cite{Kusmartsev92,Fabrizio93,Ledermann00,Dusuel03}.

\begin{figure}[tb]
    \begin{center}
      \epsfig{file=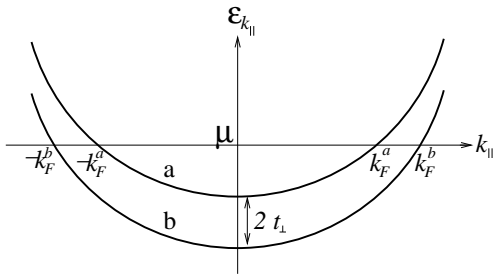,width=65mm}
    \end{center}
  \vspace{-4mm}
    \caption{%
     Energy bands of two metallic chains with bare 
     dispersions $\epsilon^a_{k_{\parallel}}$ (antibonding band) and 
     $\epsilon^b_{ k_{\parallel}}$ (bonding band).
 $\mu$ is the chemical potential.
    We consider the regime where the FS without interactions consists of four
   points $\pm k_F^a$ and $\pm k_F^b$.
 In the text $k_F^a$ and $k_F^b$ always refer to the true FS of the
interacting system, defined via Eq.~(\ref{eq:FStrue}). 
}
    \label{fig:band}
  \end{figure}

In spite of many years of research,
one fundamental aspect has only recently been addressed \cite{Dusuel03}:
the {\it{self-consistent}} renormalization of the
Fermi surface (FS). 
The band structure of
two non-interacting metallic chains consists of a  
bonding and an antibonding band, 
leading in the generic case to a FS
consisting of four points $\pm k_F^a$ and $\pm k_F^b$, as shown in
Fig.~\ref{fig:band}. 
If the density $n$ is held constant, then Luttinger's theorem
implies that the  sum $k_F^a + k_F^b = \pi n$ 
is not affected by interactions;
however, the difference $\Delta = k_F^b - k_F^a$, which
is proportional to $t_{\bot}$ in the absence of interactions,
 can be strongly renormalized.
The  FS renormalization in two coupled metallic
chains has been studied previously using
RG methods in Refs. \cite{Fabrizio93,Louis01,Dusuel03}.
However, the calculation of the FS in these works is 
either not self-consistent \cite{Fabrizio93,Louis01} in the sense
discussed in the classic book
by Nozi\`{e}res~\cite{Nozieres64},  or the self-consistency  has not been properly implemented within the
framework of the RG~\cite{Dusuel03}.
The self-consistent definition of the FS is
follows naturally from  the fact that the shape of the FS is a
characteristic property of 
a zero temperature RG fixed point~\cite{Kopietz01,Ferraz03a,Ferraz03b,Ledowski03}.
We emphasize that in our approach the concept of a flowing
FS depending on the RG flow
parameter~\cite{Fabrizio93,Louis01} never appears: 
we construct the true  FS as a
RG fixed point, which by definition
does not flow.

\begin{figure}[tb]
    \begin{center}
      \psfrag{Deltaxxx}{${\Delta} / \Delta_{1}\; \; \; $}
      \psfrag{gamma}{$\gamma$}
      \psfrag{gxxx}{${g}_0 \; \; $}
 \epsfig{file=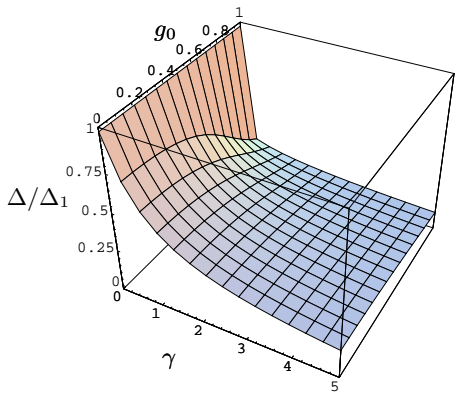,width=60mm}
\end{center}
    \vspace{-4mm}
    \caption{%
    Distance ${\Delta}  = k_F^b - k_F^a$ between the Fermi momenta
   in the Luttinger liquid phase of two spinless 
   chains coupled by weak interchain hopping $t_{\bot}$, see
    Eq.~(\ref{eq:deltares}). Here
$ \Delta_1$ is the value of $k_F^b - k_F^a$ within the
Hartree-Fock approximation,    ${g}_0$ is the
    bare dimensionless interchain backscattering interaction,
    and  $\gamma = 
     {g}_0^2 \ln ( \Lambda_0 / \Delta_{1} )^2$.   
    }
    \label{fig:FSsolution}
  \end{figure}

In this work we shall use our general 
method~\cite{Kopietz01,Ferraz03a,Ferraz03b,Ledowski03}
of obtaining the FS as a RG fixed point
to derive the self-consistency conditions for the
true Fermi momenta $k_F^a$ and $k_F^b$
of two coupled spinless chains.
In the parameter regime where the
coupled chain system is a stable Luttinger liquid,
these conditions can be cast into a simple  transcendental
equation for the distance $\Delta = k_F^b - k_F^a$ between the Fermi momenta,
which for small $t_{\bot}$ and for
weak interactions takes the form
 \begin{equation}
 \Delta =   \Delta_{1} [ 1 +   {g}_0^2 \ln ( \Lambda_0  / \Delta )^2]^{-1}
 \label{eq:deltares}
 \; .
 \end{equation}
Here $\Delta_{1}$ is the value of
$k_F^b - k_F^a$ within the Hartree-Fock approximation, 
${g}_0$ is the bare value of a suitably defined (see below)
dimensionless coupling describing interchain backscattering, and
$\Lambda_0 \gg \Delta$ is an
ultraviolet cutoff.
We emphasize that
Eq.~(\ref{eq:deltares}) depends 
only logarithmically on $\Lambda_0$, so that
the solution $\Delta$  is not very sensitive to the  precise numerical value of $\Lambda_0$.
Defining $x = \Delta_{1} / \Delta$ 
and  $\gamma = {g}_0^2 \ln ( \Lambda_0 /
\Delta_{1} )^2$, we may rewrite 
Eq.~(\ref{eq:deltares}) in the form $x = 1 + \gamma +
{g}_0^2 \ln  x^2$, from which it is easy to see graphically
that for $\gamma \gg 1$ even weak interchain backscattering
leads to a strong reduction of the distance between the two Fermi momenta, 
although for finite  $\gamma$ they never merge.
A numerical solution of Eq.~(\ref{eq:deltares}) is shown in 
Fig.~\ref{fig:FSsolution}.

Before we discuss of derivation of  Eq.~(\ref{eq:deltares}), let 
us comment on  its physical significance.
First of all,  Eq.~(\ref{eq:deltares}) predicts a strong
``attraction'' between the two Fermi momenta. 
It is tempting to extrapolate this
result to an infinite array of coupled chains: then the FS forms a continuum
consisting of two corrugated sheets.
Our two-chain result suggests that the curvature of the sheets should
decrease 
due to interactions. This in turn enforces the
nesting singularities and hence might  stabilize a Luttinger 
liquid state even in higher dimensions.
The fact that interchain back-scattering 
reduces the value of $k_F^b - k_F^a$ has already been 
noticed by Fabrizio \cite{Fabrizio93}. However,
he neither  studied  this   phenomenon quantitatively, nor
did he try to calculate the true FS self-consistently.

The renormalization of $k_F^{b} - k_F^a$ 
predicted by Eq.~(\ref{eq:deltares})
behaves rather
different from the estimate based on 
perturbation theory around the exact solution of the isolated 
chains with only forward scattering \cite{Arrigoni98,Wen90,Guinea04}.
Such an estimate leads to the 
prediction that the renormalized interchain hopping is
$t_{\bot}^{\ast} \propto t_{\bot}  | t_{\bot} / v_F \Lambda_0|^{\frac{\eta}{1- \eta}}$, 
where $\eta$ is the anomalous dimension of the Luttinger liquid
and $v_F$ the Fermi velocity
for $t_{\bot}=0$.
The fact that without interactions $k_F^b - k_F^a   \propto t_{\bot}$
suggests that in the interacting system
$k_F^{b} - k_F^a  \propto t_{\bot}^{\ast} \propto
t_{\bot} [ 1 - \eta \ln | v_F \Lambda_0 /  t_{\bot} | ]$ for small $\eta$. 
A comparison with Eq.~(\ref{eq:deltares}) shows that
this is not a valid procedure to estimate the location of the 
true FS. 


\section{Exact RG equation for the Fermi surface}

In order to derive
Eq.~(\ref{eq:deltares}) we need to keep track of the momentum- and
frequency dependence of the four-point vertex.
To do this we use here the functional 
RG method \cite{Wetterich93,Morris94,Salmhofer01,Kopietz01,Ferraz03a,Ferraz03b,Ferraz04,Busche02,Schuetz04}. 
We have derived Eq.~(\ref{eq:deltares}) using both the field
theoretical~\cite{Ferraz03a,Ferraz03b,Ferraz04} and the Wilsonian 
formulation~\cite{Kopietz01,Busche02}
of the functional RG.
In order to make contact
with previous work~\cite{Fabrizio93,Louis01}, let us outline here the derivation
within the Wilsonian approach. 
We shall briefly comment on the corresponding calculation within the field theoretical
RG in the concluding paragraph.
Starting point in the Wilsonian approach is the exact hierarchy of RG flow equations for the
irreducible self-energy
$\Sigma_{\Lambda} ( \sigma , K )$ and the one-particle irreducible
$2n$-point vertices $\Gamma^{(2n)}_{\Lambda} ( \{ \sigma_i,  K_i  \})$ 
describing the change of these quantities 
as we vary the infrared 
cutoff $\Lambda$. Here $K = ( k_{\parallel} , i \omega)$
denotes the momentum $k_{\parallel}$ along the chain and the
frequency $i \omega$. The pseudospin $\sigma = a,b$ labels the two bands.
We use the Matsubara formalism at zero temperature and the same
sign conventions for the vertices as in Ref.~[\onlinecite{Schuetz04}].
For convenience we introduce the cutoff $\Lambda$ in momentum
space, so that for a given $\Lambda$ all states with
$ | k_{\parallel} \pm k_F^{\sigma} | > \Lambda$ are integrated out.

To take the scaling properties close to the Luttinger liquid fixed point into
account, we introduce rescaled momenta $q$ and frequencies $i \epsilon$ 
by setting $k_{\parallel} =   \alpha  k_F^{\sigma} + \Lambda q$ and
$ i \omega =  v_F \Lambda  i \epsilon$, where
$\alpha = {\rm sgn} k_{\parallel} = \pm 1$ labels the right/left  half-axis.
We  define the rescaled two-point vertex by
 \begin{equation}
\tilde{\Gamma}^{(2)}_l ( \sigma , \alpha , Q ) =  \frac{Z_l^{\sigma } }{  
 v_F \Lambda }
 [ 
 \Sigma_{\Lambda} ( \sigma , K ) - \Sigma ( {\sigma}, \alpha k_F^{\sigma} ,i0)    ]
 \; , 
\label{eq:Gammadef}
 \end{equation}
which is a function of the rescaled energy-momentum
$Q = ( q , i \epsilon)$ and the logarithmic flow parameter
$ l =  \ln ( \Lambda_0 / \Lambda )$.
Here $Z_l^{\sigma}$ is the wave-function renormalization factor, and
the counter-term  
$ \Sigma ( \sigma , \alpha k_{F}^{\sigma} , i0)$
is the exact self-energy at  
the exact $k_F^{\sigma}$. 
Once we know the  counter-term
in Eq.~(\ref{eq:Gammadef}), we can calculate the
true FS for a given  chemical
potential $\mu$ from the defining equation~\cite{Nozieres64}
 \begin{equation}
 \epsilon_{ k_F^{\sigma}}^{\sigma}  + \Sigma ( {\sigma}, \alpha k_F^{\sigma}, i0 ) = \mu
 \; ,
 \label{eq:FStrue}
 \end{equation}
where $\epsilon_{ k_{\parallel}}^{\sigma}$ is the energy dispersion
of band $\sigma = a,b$.
In higher dimensions the corresponding counter-term 
$\Sigma ( {\bf{k}}_F , i0 )$ and the momenta ${\bf{k}}_F$ on the
FS can be   calculated within  
self-consistent perturbation theory in the weak coupling regime~\cite{Neumayr03}. 
On the other hand, 
in one dimension it is necessary to invoke the RG to properly treat the infrared divergencies.
A formally exact equation for the counter-term 
can be obtained from the requirement that for $ l \to \infty$
the relevant part
 \begin{equation}
{r}_l^{\sigma} \equiv \tilde{\Gamma}^{(2)}_l (  \sigma , \alpha, 0)
\end{equation}
of the two-point vertex (\ref{eq:Gammadef}) flows into a RG fixed point~\cite{Ledowski03}.
Defining the rescaled four-point vertex via
 \begin{eqnarray}
 \tilde{\Gamma}^{(4)}_l ( \sigma_1^{\prime}  \alpha_1^{\prime}   Q_1^{\prime},
\sigma_2^{\prime}  \alpha_2^{\prime}   Q_2^{\prime} ;
 \sigma_2 \alpha_2 Q_2 , \sigma_1  \alpha_1  Q_1 )  & = & ( 2 \pi v_F )^{-1}
 \nonumber 
\\
 & & \hspace{-67mm}
\times \bigl(  \prod_{i=1,2} Z_l^{\sigma_i^{\prime}} 
Z_l^{\sigma_i} \bigr)^{1/2} \Gamma^{(4)}_{\Lambda} (
   \sigma_1^{\prime}     K_1^{\prime},
\sigma_2^{\prime}   K_2^{\prime} ;
 \sigma_2 K_2 , \sigma_1 K_1 )
 \; ,
 \label{eq:Gamma4def}
 \end{eqnarray}
 the exact RG flow equation for
$\tilde{\Gamma}_l^{(2)}  ( \sigma,  \alpha , Q )   $  reads
 \begin{equation}
 \partial_l \tilde{\Gamma}^{(2)}_l =
( 1 - \eta_l^{\sigma} - Q \cdot \partial_Q ) 
 \tilde{\Gamma}^{(2)}_l  +
 \dot{\Gamma}_l^{(2)} ( \sigma  \alpha  Q )
 \; ,
 \label{eq:Gamma2flow}
 \end{equation}
where $\eta_l^{\sigma} = - \partial_l \ln Z_l^{\sigma}$ is the flowing 
anomalous dimension, $Q \cdot \partial_Q = q \partial_q + \epsilon \partial_{\epsilon}$, 
and the inhomogeneity is
 \begin{eqnarray}
 \dot{\Gamma}_l^{(2)} ( \sigma  \alpha  Q ) & = & -
 \sum_{\sigma^{\prime}  \alpha^{\prime} } 
 \int \frac{dq^{\prime} d \epsilon^{\prime}}{ ( 2 \pi)^2}
 \dot{G}_l ( \sigma^{\prime}  \alpha^{\prime}  Q^{\prime} )
 \nonumber
 \\
 &  & \hspace{-10mm} \times
 \tilde{\Gamma}_l^{(4)} ( \sigma \alpha Q, \sigma^{\prime} \alpha^{\prime} Q^{\prime} ;
 \sigma^{\prime} \alpha^{\prime} Q^{\prime} , \sigma \alpha Q )
 \; .
 \label{eq:dotGamma}
 \end{eqnarray}
Here for sharp cutoff and linear dispersion
 \begin{equation}
 \dot{G}_l ( \sigma  \alpha  Q ) = - \delta ( 1 - | q | ) / [ Z_l^{\sigma} ( i 
 \epsilon - \alpha {v}^{\sigma}_0 q ) - 
\tilde{\Gamma}^{(2)}_l ( \sigma \alpha Q )]
 \; ,
 \end{equation}
with the dimensionless Fermi velocities
${v}^{\sigma}_0 = v_F^{\sigma} / v_F$, 
where $v_F^{\sigma}$ is the bare
Fermi velocity associated with band $\sigma$.
Because the coupling $r_l^{\sigma}$ is relevant with scaling dimension
$+1$, its  initial value ${r}_0^{\sigma}$
must to be fine tuned in order to approach a finite limit
for $l \to \infty$. This leads to the condition~\cite{Ledowski03}
 \begin{equation}
 {r}_0^{\sigma} = - \int_0^{\infty} dl e^{ - l + \int_0^{l} dt \eta_t^{\sigma} }
   \dot{\Gamma}_l^{(2)} ( \sigma , \alpha , 0 )
 \; ,
 \label{eq:r0}
 \end{equation}
which relates
the inital value  ${r}_0^{\sigma}$ to the 
RG flow  on the entire RG trajectory. Assuming that
$\Lambda_0$ is sufficiently large so that
$\Sigma_{\Lambda_0}$  can be neglected and
$Z_0^{\sigma} \approx 1$, we obtain from  Eq.~(\ref{eq:Gammadef}) for the counter-term
 \begin{equation}
\Sigma ( {\sigma} , \alpha k_F^{\sigma} , i0 ) = - v_F \Lambda_0 {r}_0^{\sigma}
 \; .
\end{equation}
At constant density $n = (k_F^a + k_F^b ) / \pi$ we then find
 \begin{equation}
 k_F^b - k_F^a \equiv \Delta = \Delta_0 +  2 \Lambda_0 ( {v}_0^{b} + {v_0^a} )^{-1}
 ({r}_0^b - {r}_0^a )
 \; ,
 \label{eq:FSshift}
 \end{equation}
where $\Delta_0$ is the value of $k_F^b - k_F^a$ 
at the same density but without interactions.

\section{Weak coupling truncation for the  effective  interaction and  the self-energy}

The formally exact Eqs.~(\ref{eq:Gamma2flow}--\ref{eq:FSshift}) are now our 
starting point to calculate 
$\Delta$ self-consistently. 
Note that the concept of a ``flowing FS'' never appears in our 
approach; by definition, $k_F^a$ and $k_F^b$ are the true Fermi momenta of the interacting system,  
to be determined self-consistently from
Eqs.~(\ref{eq:r0}) and (\ref{eq:FSshift}). To  make further progress, we need 
an approximate expression for the effective interaction
$\tilde{\Gamma}^{(4)}_l$ in Eq.~(\ref{eq:dotGamma}).
To simplify the analysis, we neglect vertices describing
intrachain umklapp scattering as well as chiral vertices, 
involving particles moving in the same direction.
The latter 
are expected to give finite renormalizations of the 
Fermi velocities, which we ignore here. 
Totally we should then keep track of the RG flow
of the following five vertex functions,
 \begin{subequations}
\begin{eqnarray}
   \tilde{f}_l^{bb} ( Q_1^{\prime} , Q_2^{\prime} ; Q_2 , Q_1 )
   =  
\tilde{\Gamma}^{(4)}_l ( b Q_1^{\prime}, b Q_2^{\prime}; 
 b Q_2, b Q_1 )  , 
 & &
 \label{eq:fbb}
 \\
 \tilde{f}_l^{aa} ( Q_1^{\prime} , Q_2^{\prime} ; Q_2 , Q_1 )
  =  
\tilde{\Gamma}^{(4)}_l ( a Q_1^{\prime}, a Q_2^{\prime}; 
 a Q_2, a Q_1 )
  , & &
 \label{eq:faa}
 \\
 \tilde{f}_l^{ab} ( Q_1^{\prime} , Q_2^{\prime} ; Q_2 , Q_1 )
  = 
\tilde{\Gamma}^{(4)}_l ( a Q_1^{\prime}, b Q_2^{\prime}; 
 b Q_2, a Q_1 )
  , & &
 \label{eq:fab}
 \\
\tilde{u}_l ( Q_1^{\prime} , Q_2^{\prime} ; Q_2 , Q_1 )
  = 
\tilde{\Gamma}^{(4)}_l ( b Q_1^{\prime}, b Q_2^{\prime}; 
 a Q_2, a Q_1 )  ,
 & & 
 \label{eq:u}
 \\
\tilde{g}_l ( Q_1^{\prime} , Q_2^{\prime} ; Q_2 , Q_1 )
  = 
\tilde{\Gamma}^{(4)}_l ( b Q_1^{\prime}, a Q_2^{\prime}; 
 b Q_2, a Q_1 )  ,
 & & 
 \label{eq:g}
 \end{eqnarray}
 \end{subequations}   
where on the right-hand side the direction labels
are  $ (\alpha_1^{\prime} , \alpha_2^{\prime} ; \alpha_2,  \alpha_1)
 =  ( +,-;-,+)$. 
Here $\tilde{f}_l^{\sigma \sigma^{\prime}}$ describes  forward scattering 
of two fermions associated with bands $\sigma$ and $\sigma^{\prime}$,
$\tilde{u}_l$ describes
interband umklapp scattering (also called pair tunneling), and
$\tilde{g}_l$ corresponds to interband backscattering. 
The functional RG equations for these five functions follow from the
general flow equations given in Ref.~\cite{Kopietz01} in a straightforward
way. If we set 
$Q_i = Q_i^{\prime} =0$ and denote the corresponding coupling
{\it{constants}} by
$f^{\sigma \sigma^{\prime}}_l$, $u_l$ and $g_l$ (without a tilde),
we obtain the one-loop RG flow equations 
 \begin{equation}
 \partial_l u_l = - u_l f_l  \;,  \;  \partial_l g_l = g_l f_l
  \; ,  \; \partial_l f_l =   - A u_l^2 + B_l g_l^2
 \label{eq:ugflow}
 \; ,
 \end{equation}
where 
 \begin{equation}
f_l = f^{aa}_l / {v}_0^a + f^{bb}_l / {v}_0^b - 2
 f^{ab}_l /  \bar{v}_0
 \; ,
 \end{equation}
with
$A = 2 /(  {v}_0^a  
{v}_0^b) +   2 / \bar{v}_0^2  $ and the average dimensionless velocity
$\bar{v}_0 = ( {v}^a_0 + {v}^b_0)/2$.
The function 
 \begin{equation}
B_l =  2 C_l /(  {v}_0^a  
{v}_0^b) +   2 \bar{C}_l / \bar{v}_0^2
 \end{equation}
depends 
on the rescaled Fermi point difference
$\tilde{\Delta}_l = \Delta e^l / \Lambda_0$ 
via
 \begin{eqnarray}
 C_l & = & \Theta ( e^l -1 - 2 | \tilde{\Delta}_l |) / ({ 1 + 
 | \tilde{\Delta}_l |})
 \; ,
 \\
 \bar{C}_l  & = & \Theta ( e^l -1 - 2 | \tilde{\Delta}_l | )
 \frac{1}{2}  \sum_{\sigma}  \frac{\bar{v}_0 }{ 
 \bar{v}_0 + {v}_0^{\sigma} | \tilde{\Delta}_l |   }
 \; .
\end{eqnarray}
The RG equations (\ref{eq:ugflow}) are equivalent to those derived by
Fabrizio \cite{Fabrizio93} at one-loop order. 
Note that the RG flow of $u_l$ and $g_l$ 
couples to the forward scattering interactions only via
the combination $f_l$ defined above.
The flow of $f_l^{\sigma \sigma^{\prime}}$  is
\begin{eqnarray}
 {v}_0^{a} \partial_l f_l^{bb}  & = &  
- u_l^2 + C_l g_l^2  = {v}_0^b \partial_l f_l^{aa} 
 \; ,
 \\
  \bar{v}_0 \partial_l f_l^{ab}  & = &   
u_l^2 - \bar{C}_l g_l^2
\; .
 \end{eqnarray}
As discussed by Fabrizio \cite{Fabrizio93},
the above RG equations predict 
a finite regime of initial values
$f^{\sigma \sigma^{\prime}}_0$, $u_0$, and $g_0$ where
the system is a stable Luttinger liquid, characterized by
finite forward scattering 
couplings $f_l^{\sigma \sigma^{\prime}}$ and vanishing
$u_l$ and $g_l$ for $l \to \infty$.
%
%
%
%
%

We now calculate the true FS in the Luttinger liquid regime using our general
Eqs.~(\ref{eq:Gamma2flow}--\ref{eq:FSshift}). 
Assuming that
$f_l^{\sigma \sigma^{\prime}}$, $u_l$, and $g_l$ are small compared with
unity, we may adopt the same strategy
as in Refs.\cite{Busche02,Ledowski04}: we expand the
right-hand side of the flow equations for the 
momentum- and frequency dependent vertices
defined in Eqs.~(\ref{eq:fbb}--\ref{eq:g}) to second
order in $f_l^{\sigma \sigma^{\prime}}$, $u_l$, and $g_l$.
The flow of the above vertex functions is then given by the following set of equations,
 \begin{subequations}
\begin{eqnarray}
  D_l^{bb} \tilde{f}_l^{bb} ( Q_1^{\prime} , Q_2^{\prime} ; Q_2, Q_1 )
  & = & 
\nonumber
 \\
& & \hspace{-42mm}
( f_l^{bb})^2 [ \dot{\chi}_l^{bb} ( Q_1 - Q_2^{\prime} ) -  \dot{\chi}_l^{bb} ( Q_1 + Q_2  ) ]
 \nonumber
 \\
 &  & \hspace{-42mm} - u_l^2  \dot{\chi}_l^{aa} ( Q_1 + Q_2 ) + g_l^2  
\dot{\chi}_l^{aa} ( Q_1 - Q_2^{\prime} + 2 \tilde{\Delta}_l ) \; ,
 \label{eq:flow1}
\end{eqnarray}

\begin{eqnarray}
 D_l^{aa} \tilde{f}_l^{aa} ( Q_1^{\prime} , Q_2^{\prime} ; Q_2, Q_1 )
  & =  & 
\nonumber
 \\
& & \hspace{-42mm}
( f_l^{aa})^2 [ \dot{\chi}_l^{aa} ( Q_1 - Q_2^{\prime} ) -  \dot{\chi}_l^{aa} ( Q_1 + Q_2  ) ]
 \nonumber
 \\
 &  & \hspace{-42mm} - u_l^2  \dot{\chi}_l^{bb} ( Q_1 + Q_2 ) + g_l^2  
\dot{\chi}_l^{bb} ( Q_1 - Q_2^{\prime} - 2 \tilde{\Delta}_l ) \; ,
 \label{eq:flow2}
\end{eqnarray}

\begin{eqnarray}
 D_l^{ab} \tilde{f}_l^{ab} ( Q_1^{\prime} , Q_2^{\prime} ; Q_2, Q_1 )
  & =  & 
\nonumber
 \\
& & \hspace{-42mm}
( f_l^{ab})^2 [ \dot{\chi}_l^{ab} ( Q_1 - Q_2^{\prime} ) -  \dot{\chi}_l^{ab} ( Q_1 + Q_2  ) ]
 \nonumber
 \\
 &  & \hspace{-42mm} + u_l^2  \dot{\chi}_l^{ba} ( Q_1 - Q_2^{\prime} ) - g_l^2  
\dot{\chi}_l^{ba} ( Q_1 + Q_2 - 2 \tilde{\Delta}_l ) \; ,
 \label{eq:flow3}
\end{eqnarray}

\begin{eqnarray}
 D_l^{ab} \tilde{u}_l ( Q_1^{\prime} , Q_2^{\prime} ; Q_2, Q_1 )
  & =  & 
\nonumber
 \\
& & \hspace{-42mm}
 f_l^{ab} u_l [ \dot{\chi}_l^{ab} ( Q_1 - Q_2^{\prime} ) +  \dot{\chi}_l^{ba} ( Q_1 - Q_2^{\prime}  ) ]
 \nonumber
 \\
 &  & \hspace{-42mm} - f^{bb}_l u_l  \dot{\chi}_l^{bb} ( Q_1 + Q_2 ) - f_l^{aa} u_l  
\dot{\chi}_l^{aa} ( Q_1 + Q_2  ) \; ,
\label{eq:flow4}
\end{eqnarray}

\begin{eqnarray}
 D_l^{ab} \tilde{g}_l ( Q_1^{\prime} , Q_2^{\prime} ; Q_2, Q_1 )
  & =  & 
\nonumber
 \\
& & \hspace{-42mm}
 - f_l^{ab} g_l [ \dot{\chi}_l^{ab} ( Q_1 + Q_2 ) +  \dot{\chi}_l^{ba} ( Q_1 +  Q_2  ) ]
 \nonumber
 \\
 &  & \hspace{-42mm} + f^{bb}_l g_l  \dot{\chi}_l^{bb} ( Q_1 - Q_2^{\prime} ) + f_l^{aa} g_l  
\dot{\chi}_l^{aa} ( Q_1 - Q_2^{\prime}  ) \; .
 \label{eq:flow5}
\end{eqnarray}
\end{subequations}
Here $Q + \tilde{\Delta}_l $ is a short notation for $ ( q + \tilde{\Delta}_l , i \epsilon )$, the generalized
susceptibilities   $\dot{\chi}_l^{\alpha \beta} ( Q )$ 
are given by
 \begin{eqnarray}
 \dot{\chi}_l^{\alpha \beta} ( Q ) & = & 
\frac{ \Theta ( e^l -1 - | q | )    }{ 2 \bar{v}_0  + v_0^{\alpha} | q | - i \epsilon {\rm sgn} q }
 \nonumber
 \\
 &  + &
\frac{ \Theta ( e^l -1 - | q | )   }{ 2 \bar{v}_0  + v_0^{\beta} | q | + i \epsilon {\rm sgn} q }
 \; ,
 \end{eqnarray}
where  $\alpha , \beta \in \{ a,b \}$,   and
 \begin{equation}
 D^{\alpha \beta}_l = \partial_l + \eta_l^{\alpha} + \eta_l^{\beta}
 + \sum_{i=1}^2  ( Q_i \cdot \partial_{Q_i}  + Q_i^{\prime} \cdot \partial_{ Q_i^{\prime}} )
 \end{equation}
 are comoving derivatives.
The linear partial differential equations 
(\ref{eq:flow1}--\ref{eq:flow5})
can easily be solved exactly \cite{Busche02}, so that we can obtain an explicit 
expression
for the flowing momentum- and frequency dependent effective interaction 
 $
 \tilde{\Gamma}_l^{(4)} ( \sigma \alpha Q, \sigma^{\prime} \alpha^{\prime} Q^{\prime} ;
 \sigma^{\prime} \alpha^{\prime} Q^{\prime} , \sigma \alpha Q )$.
Substituting this expression
into Eq.~(\ref{eq:dotGamma}) and performing the $Q^{\prime}$-integration we find
 \begin{eqnarray}
 \dot{\Gamma}^{(2)}_l ( a,+, Q ) & = & f^{aa}_l + f_l^{ab}
 - \partial_l [  f^{aa}_l + f_l^{ab} ]
\nonumber
 \\ &  & \hspace{-20mm} 
 - 2 (f_l^{aa})^2 I_{a,l}^{aa} ( Q ) - 2 (f_l^{ab})^2 I_{b,l}^{ab} ( Q )
 \nonumber
 \\ &  & \hspace{-20mm} 
 - u_l^2 [ I_{a,l}^{bb} ( Q ) + I_{b,l}^{ba} ( Q ) ]
 \nonumber
 \\
 & & \hspace{-20mm} - g_l^2 [ I_{a,l}^{ bb} ( q - 2 \tilde{\Delta}_l , i 
 \epsilon ) + I_{b,l}^{ba} ( q - 2 \tilde{\Delta}_l , i \epsilon ) ]
\; ,
 \label{eq:dotgammares}
 \end{eqnarray}
where  for $\alpha , \beta , \gamma \in \{ a,b \}$
 the integrals $I^{ \alpha \beta }_{ \gamma ,l} ( Q )$ are
 \begin{eqnarray}
 I^{\alpha \beta }_{ \gamma , l} ( Q ) & = & \frac{1}{ {v}^{\alpha}_0 
 + {v}^{\beta}_0 } \sum_{ n = \pm 1}
 n \Theta ( e^{l} -1 - | q + n | )
 \nonumber
 \\
 &  & \hspace{-20mm} \times \Biggl\{
 \Theta ( n s_{ q +n} ) 
 \ln \left[ \frac{ e^l ( {v}^{\alpha}_0 + {v}^{\beta}_0 )
 + {v}^{\gamma}_0 - {v}^{\beta}_0 | q +n | - i n \epsilon }{
  {v}^{\alpha}_0 + {v}^{\beta}_0 
 + {v}^{\gamma}_0 + {v}^{\alpha}_0 | q +n | - i n \epsilon}
  \right]
 \nonumber
 \\
 & &  \hspace{-20mm} +
\Theta ( - n s_{ q +n} ) 
 \ln \left[ \frac{ e^l ( {v}^{\alpha}_0 + {v}^{\beta}_0 )
 + {v}^{\gamma}_0 - {v}^{\alpha}_0 | q +n | - i n \epsilon }{
  {v}^{\alpha}_0 + {v}^{\beta}_0 
 + {v}^{\gamma}_0 + {v}^{\beta}_0 | q +n | - i n \epsilon}
  \right]
 \Biggr\}
 \,,
 \nonumber
 \\
 \label{eq:Idef}
\end{eqnarray}
with $s_{ q +n } = {\rm sgn} ( q + n )$.
The function
$ \dot{\Gamma}^{(2)}_l ( b,+, Q )$ can be obtained
from Eq.~(\ref{eq:dotgammares})
by replacing $ a \leftrightarrow b$  
and $\tilde{\Delta}_l \to
 - \tilde{\Delta}_l$ on the right-hand side.
Once we know the function $\dot{\Gamma}^{(2)}_l ( \sigma ,+, Q )$, we may
calculate the
flowing anomalous dimension from~\cite{Busche02}
 \begin{equation}
\eta^{\sigma}_l = - \partial  \dot{\Gamma}^{(2)}_l ( \sigma ,+, Q ) /
\partial ( i \epsilon ) |_{ Q = 0}
 \;  .
 \end{equation}

We can reproduce the result by
Louis {\it{et al.}}~\cite{Louis01} if we simply set
 $\dot{\Gamma}^{(2)}_l ( a,+, Q )  \approx  f^{aa}_l + f_l^{ab}$
instead of Eq.~(\ref{eq:dotgammares}), which amounts to 
replacing the effective interaction
 $
 \tilde{\Gamma}_l^{(4)} ( \sigma \alpha Q, \sigma^{\prime} \alpha^{\prime} Q^{\prime} ;
 \sigma^{\prime} \alpha^{\prime} Q^{\prime} , \sigma \alpha Q )$
in Eq.~(\ref{eq:dotGamma}) by its value at $Q = Q^{\prime} =0$.
This approximation does not consistently take into account
all contributions to $r_0^{\sigma}$ 
which are quadratic in the bare couplings,
because  
such second order terms
are also generated by
the $Q^{\prime}$-dependence of the four-point vertex in Eq.~(\ref{eq:dotGamma}).

To calculate the FS from Eq.~(\ref{eq:FSshift}), we still need to perform the
trajectory integral (\ref{eq:r0}). Since we are interested in the leading
behavior for small $t_{\bot}$, we may at this point 
ignore the small difference between the bare Fermi velocities,
setting ${v}^a_0 = {v}^b_0 = \bar{v}_0 = 1$. To second 
order in the couplings, we may also set $\eta_t^{\sigma} =0$  
in Eq.~(\ref{eq:r0}). With these approximations the
integral obtained by  substituting 
Eq. ~(\ref{eq:dotgammares}) into Eq.~(\ref{eq:r0}) can be
performed analytically to second order
in the bare couplings.  For $\tilde{\Delta} \equiv 
\Delta / \Lambda_0 \ll 1$ the result is 
 $r_0^a =   - f_0^{aa}  -f_0^{ab} - g_0^2 \tilde{\Delta}
\ln | \tilde{\Delta}|$ and
 $r_0^b =   - f_0^{bb}  -f_0^{ba} + g_0^2 \tilde{\Delta}
\ln |\tilde{\Delta}|$  to leading
logarithmic order.
Substituting 
this into Eq.~(\ref{eq:FSshift}) we arrive at
Eq.~(\ref{eq:deltares}), with
$\Delta_{1} = \Delta_0 - \Lambda_0 ( f^{bb}_0 - f^{aa}_0 )$, where
we have used $f_0^{ab} = f_0^{ba}$.

Finally, let us point out that we have also derived Eq.~(\ref{eq:deltares}) within the
field-theoretical RG, where we  calculate the renormalized vertex functions
directly from perturbation theory.  Setting  again $v_0^a = v_0^b = 1$ we obtain for
the difference between the self-energies 
$\Sigma_{\sigma} = \Sigma ( \sigma , k_F^{\sigma} , i 0 )$
at the true FS, up to two-loop order,
\begin{equation}
\Sigma_b - \Sigma_a = - 2 t_{\bot} + v_F \Lambda_0 [ f_R^{bb} - f_R^{aa}  -  g_R^2 \tilde{\Delta}
 \ln  \tilde{\Delta} ^2 ]
 \; ,
 \end{equation}
where $f_R^{\sigma \sigma^{\prime} }$ and $g_R$ are
the renormalized couplings.  Since up to this order we do not distinguish
between bare and renormalized couplings, 
the combination of this result with the definition (\ref{eq:FStrue}) 
immediately leads to Eq.~(\ref{eq:deltares}).


\section{Summary and conclusions}

In summary, we have presented a fully self-consistent calculation of the
true FS of two coupled metallic chains in the regime where
the Luttinger liquid fixed point is stable.
Our final result for the renormalized FS
is given by the solution of the transcendental  
equation  (\ref{eq:deltares}) shown in Fig.~\ref{fig:FSsolution}, which
is perhaps the simplest example for an explicit solution of the
self-consistency problem for the true FS discussed by Nozi\`{e}res \cite{Nozieres64}.
We find that  for
small interchain hopping $t_{\bot}$  even 
weak interchain backscattering can lead to a strong 
reduction of the distance between the Fermi points
associated with the bonding and the antibonding band. 
It is tempting to speculate that in higher dimensions a similar
smoothing effect of a corrugated FS 
stabilizes a strongly correlated non-Fermi liquid state.
Our method is  general and can be used to
calculate the true FS in higher dimensions  within the
framework of the RG; of course, in this case  the resulting
integral equations can only be solved numerically.


This work was completed while P.K. was visiting the
{\it{International Center of Condensed Matter Physics (ICCMP)}}
at the University of Bras{\'\i}{}lia, Brazil. 
He would like to thank the ICCMP for its  hospitality and
for financial support.

  
\end{document}